# Exploration and Practice of Improving Programming Ability for the Undergraduates Majoring in Computer Science

Guowu Yuan, Shicai Liu

School of Information Science & Engineering, Yunnan University, Kunming 650504, China

**Abstract:** Programming ability is one of the most important abilities for the undergraduates majoring in computer science. Taking Yunnan University as an example, the necessity and importance of improving the ability of programming is analyzed in this paper. The exploration and practice of improving students' ability of programming are discussed from four aspects: arrangement and reform of programming curriculums, construction of online programming practice innovation platform, certification of programming ability and organization of programming competitions. These reforms have achieved good results in recent years, which can provide reference for the practical teaching reform of computer specialty in relevant universities.

## 0 Introduction

Computer science and technology is a practical engineering major, and the programming ability of graduates is not only the most important indicator for measuring the quality of graduates, but also the most important skill that affects the future development of students [1,2]. These students need to master the basic knowledge of program design, data structure and algorithm, be able to select the appropriate programming language, accurately and skillfully complete the programming and debugging work for a given problem, and achieve the expected goal of the program. The computer major is one of the most important majors in the construction of "New Engineering". With the continuous advancement of the national strategy of artificial intelligence and the "Internet+" action plan, and the widespread application of data-intensive scientific research paradigms, the national governance, scientific research, technological research and development, and public life are almost inseparable from the strong support of computer technology.

However, the current assessment of programming-related curriculums is mainly as written examinations. The traditional computer technology and software professional and technical qualification examinations conducted by the Ministry of Human Resources and Social Security and the Ministry of Industry and Information Technology [3] (originally called China's computer software professional and technical qualifications examinations) are also mainly written examinations, and it focus on the basic concepts and the examination of certain knowledge points that is suitable for written examination. Therefore, the examinee may lack the ability of programming and debugging on the computer, which results in high scores and low abilities, and good scores on the paper, but the actual programming ability on the computer is very poor. In addition, the subjects in the graduate entrance examination, such as data structure, programming, algorithm design and analysis, are mostly written tests. Although the programming practice is added in the second round of postgraduate in some universities, it is still difficult to select the candidates with excellent programming ability for the computer major in most universities. Part of the candidates even gave up the practice of computer programming in junior and senior high school, and put all their energy into the review of the knowledge points of postgraduate subjects.

These candidates have high scores in the initial examination, but poor programming ability. These candidates do not have the engineering literacy of computer major, and lack the logic thinking ability of programming, which brings great difficulties to the subsequent training.

In addition, we got the data from the professional catalogue from the Ministry of Education[4] and the statistics from Professor Zongli Jiang who is the vice chairman of the computer teaching steering committee of the Ministry of Education. In 2019, there are 3,736 computer majors with major codes beginning with 0809 across the country, and about 1.5 million students who are studying in the campus. Among all majors in the country, the number of students majoring in computer is the largest. Table 1 shows the specific number of each major. According to the statistics from the National Bureau of Statistics, the average salary of the IT industry was RMB 130,366 in 2017 and RMB 141,962 in 2018, and the graduates' income has been at the top of the list for many years.

However, the overall employment rate of computer majors is not in an absolutely leading position, and in some provinces, computer majors even have received a yellow card. On the one hand, the students majoring in computer have high expectations for employment; On the other hand, many enterprises report that they cannot find satisfactory graduates of computer major. The questionnaire survey of employment units shows that the graduates are mainly lack of practical ability, programming ability, which cannot meet the requirements of employers.

**Table 1** Statistics on the number of computer majors in China in 2019

| Major Code | Major Name | Number of Major |
|---|---|---|
| 080901 | Computer Science and Technology (Engineering / Science) | 1002 |
| 080902 | Software Engineering | 632 |
| 080903 | Network Engineering | 429 |
| 080904K | Information Security (Engineering / Science / Management) | 132 |
| 080905 | IoT Engineering | 546 |
| 080906 | Digital Media Technology | 270 |
| 080907T | Intelligent Science and Technology | 153 |
| 080908T | Spatial Information and Digital Technology | 21 |
| 080909T | Electronics and Computer Engineering | 9 |
| 080910T | Data Science and Big Technology | 481 |
| 080911TK | Cyberspace Security | 53 |
| 080912T | New Media Technology | 3 |
| 080913T | Film Production | 2 |
| 080914TK | Security Technology | 3 |
| **0809** | **Computers (Total)** | **3736** |

Therefore, how to improve the practical ability of programming has become a key problem. The training of programming ability is a gradual process. The computer science and technology of Yunnan university has done relevant work from the following four aspects: program design curriculum construction, online programming practice platform construction, program design ability certification and program design competition: (1) In the arrangement of programming curriculums, we ensure that programming is uninterrupted from the freshman to juniors; (2) We build an online programming practice innovation platform based on GitLab; (3) In the programming ability certification, we ensure that each student can achieve a certain score on the Certified Software Professional of China Computer Federation (CCF CSP); (4) In the programming competitions, most undergraduates are encouraged to participate in at least one programming competition. From the four levels of curriculum, platform, certification, and competition, we have built a "four-dimensional" and "Positive Feedback" computer programming curriculum system for Yunnan University. We extends the classroom teaching chain of programming curriculums, expands the programming ability promotion channels, and consolidates the basis of cultivating the ability of programming. From the perspective of the implementation effect in recent years, these methods have achieved good results.

# 1 Construction and Reform of Programming Curriculums

## 1.1 Arrange multi-dimensional programming curriculums to expand the teaching chain

In the revision of training program for computer science and technology, relevant courses of programming can be divided into three categories: required course, elective courses and comprehensive practice. Each student needs to complete required course and comprehensive practice, but can choose the different elective courses which must meet the minimum credit requirements of elective course. The three types of courses are shown in Table 2.

Table 2 Programming Curriculum Arrangements in the Training Program

| Semester | Required Course | Elective Course | Comprehensive Practice |
|---|---|---|---|
| 1 | Computer Programming, Computer Programming Experiments | | |
| 2 | Data Structure, Data Structure Experiment | | |
| 3 | Object-oriented Programming and Training, Computer Graphics Experiments, Algorithm Design and Analysis Experiments | Numeral Calculations Experiment | |
| 4 | Computer Network Experiment | Mathematical Modeling and Experiments, WEB Application Software Development Training, Digital Image Processing Experiments | |
| 5 | Database Technology Experiment, Operating System Experiment | Assembly Language Programming | Scientific Research Training |
| 6 | Compilation Technology Experiment, Software Engineering Practice | Big Data Basic Experiments, Artificial Intelligence Experiments | Innovation Experiment |
| 7 | | Cloud Computing Experiment | Professional Internship, Programming Ability Test |
| 8 | | | Graduation Internship, Graduation Design |

In the first three years, we guarantee at least one compulsory programming curriculum every semester, so that students can program continuously. From the third semester, elective courses of programming are gradually added, and students choose according to their own interests. From the 5th semester, comprehensive practice curriculums will be added. Scientific research training, innovation experiments, graduation practice and graduation design will improve students' comprehensive ability, and these practice require programming. Some IT enterprise courses are arranged in professional internships. Some IT enterprises are introduced to schools to concentrate on professional internships in the first two weeks of summer vacation after the end of the sixth semester final exam. In the past two years, the basic curriculums of Huawei, such as cloud computing, big data, routing and switching, have been implemented. Huawei' engineers arrived at the school for two weeks to conduct on-campus professional internship training for the students with 80 hours. To obtain the credit of programming ability test, each student should take part in the CCF CSP (the Certified Software Professional of China Computer Federation), and achieve the basic qualification line, or pass China's computer software professional and technical qualifications examinations.

The arrangement of these curriculums guarantees that students will need to continue programming training and improve their programming skills. After obtaining the credit of programming ability test, students are required to achieve a certain programming ability with the third-party standard.

### 1.2 Reform the assessment method of programming curriculums

In the final examination of the previous course, the proportion of paper-based examination results in the total score is too high, while the proportion of experiment is too low. Now, we have adjusted the assessment method, the main measures are as follows:

(1) Part of the experiment curriculums are separated from the theoretical curriculums into separate experimental curriculums, which are fully assessed according to the practical situation. These curriculums include computer programming experiments, data structure experiments, computer graphics experiments, database technology experiments, operating system experiments, software engineering internships, etc.;

(2) The course can be applied as an examination reform course. The paper examination will be cancelled and the proportion of experiment courses will be increased. These courses include mathematical modeling and experiment, web application software development training, big data foundation, introduction to artificial intelligence, etc.

## 2 Construction of Online Programming Practice Innovation Platform

In the teaching of programming experiments, the teachers manually check the students' code one by one, which is heavy and inefficient, and it is impossible to avoid code plagiarism. As a result, sometimes there are only a few versions of the programming codes in the class. Some students copy the codes from the outstanding students to meet the teacher's inspection, which seriously harms the students' hands-on ability development.

It's a good way to check the code and score automatically. On the Internet, there are already some free programming online test systems. However, due to the factors such as system access, resource management mode, security and privacy protection, such systems still cannot be effectively used in teaching and play the expected role. Therefore, we have built our own online teaching platform for programming courses, which has both the function of teaching organization management and the module of continuous integration of open source programs.

### 2.1 System architecture and function module design

Gitlab [5] is an open-source version management system, which realizes the self-managed Git project warehouse. It can access public or private projects through the web interface. It can browse the source code, manage defects and comments, and access the warehouse by the management team. It is easy to browse the submitted version and provide a file history library. GitLab is an open-source program continuous integration framework widely used in the world. Its program version control, multi-person collaborative development, code snippet collection and reuse provide a good reference and platform foundation for the construction of practice and innovation platform of programming courses. It also provide a necessary tool for modern computer talents cultivation, such as open-source program design, continuous iterative integration, resource cloud. Therefore, we try to introduce the industry's open source programming platform into the practice of programming course teaching, integrate the above-mentioned computer training and curriculum reform ideas, and build an online programming practice innovation platform based on Gitlab.

According to our technical requirements, we use mature and controllable open source projects for system integration, involving four main functional modules: Code hosting, version control, code automatic execution and code online editing. On the highly available private cloud platform of the school virtual data center, the virtual server is created in the way of virtual machine, and Gitlab Community Edition is deployed as the business core of the whole system. The system architecture as shown in Figure 1 is proposed. The traditional centralized storage structure is adopted in the data center, the computing function is carried out on the independent computing server, the data is uniformly stored in the disk array of the storage system, and the internal storage network of the data center is built in the way of San; the continuous integration function of Gitlab is realized by the technology of runner. Each Gitlab runner is a daemons running on the continuous integration server, which maintains a heartbeat connection to the Gitlab master server through the HTTP channel.

The execution process of a continuous integration is described as follows: when there is a new continuous integration task, the Gitlab master server issues an instruction to the registered Gitlab executor; and the Gitlab executor issues a request to the code hosting system of the Gitlab master server according to the instruction, the project code of this task is downloads through the Git protocol, and then executes the project according to the continuous integration script defined in the project code. In the execution process, the Gitlab executor sends the intermediate results back to the task scheduling system of the Gitlab main server in real time. After execution, the Gitlab executor uploads the generated product (artifact) to the code hosting system of the main server, and finally destroys the copy of the local project.

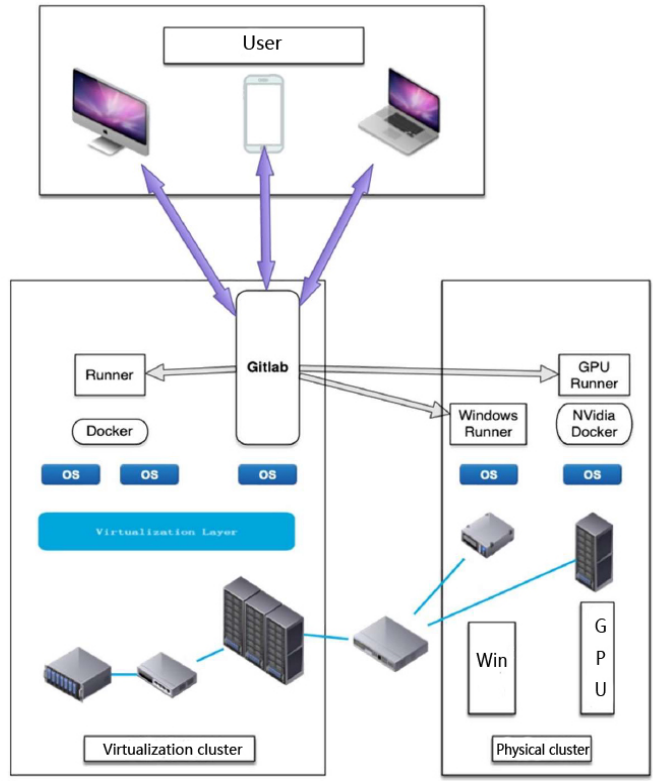

Figure 1 system architecture of the platform

2.2 Key technologies of platform construction

The following describes the key technologies of platform, focusing on the deployment and configuration of Gitlab executor and the use of Docker container.

2.2.1 Deployment and configuration of Gitlab executor

Gitlab executor is installed in Linux, MacOS, windows, FreeBSD and other host operating systems as a daemon. It has a comprehensive support for various operating systems, supports the installation of Docker services to independent Docker hosts or Kubernets clusters, and also supports the horizontal automatic scaling function based on Docker machine technology. The tip of Gitlab installation is shown in Figure 2.

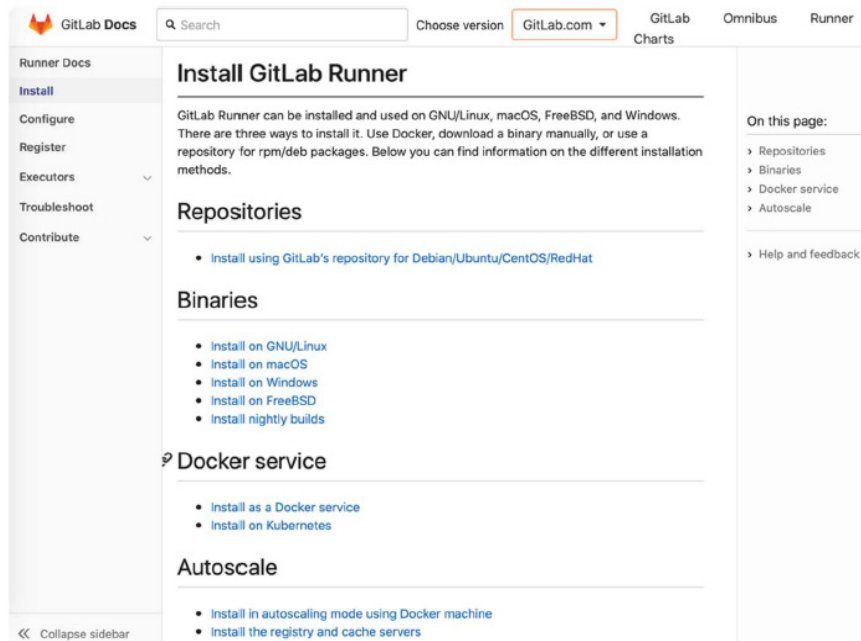

Figure 2 Gitlab actuator installation information

2.2.2 Docker container

On the continuous integration server, Gitlab executors mainly include 7 types of executors running project code, including SSH executors, VirtualBox executors, Docker executors, Docker-ssh executors, parallels executors, Shell executors, and custom executors. In our platform, we mainly use Docker executor and Shell executor. Docker executor calls the Docker container management process on the host where Gitlab executor is located, and starts a Docker container on the host to execute project code according to the Docker image requirements defined in the continuous integration script. Shell executor directly calls Shell of the host where Gitlab executor is located to execute project code. Shell executor is mainly used to provide the code execution environment of Windows system.

The continuous integration script shown in Figure 3 can execute python, Java and C code in a project, and the execution result is shown in Figure 4.

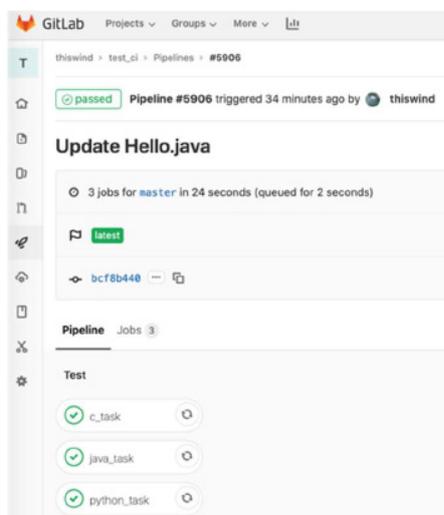

Figure 3 Code written in Python, Java, and C

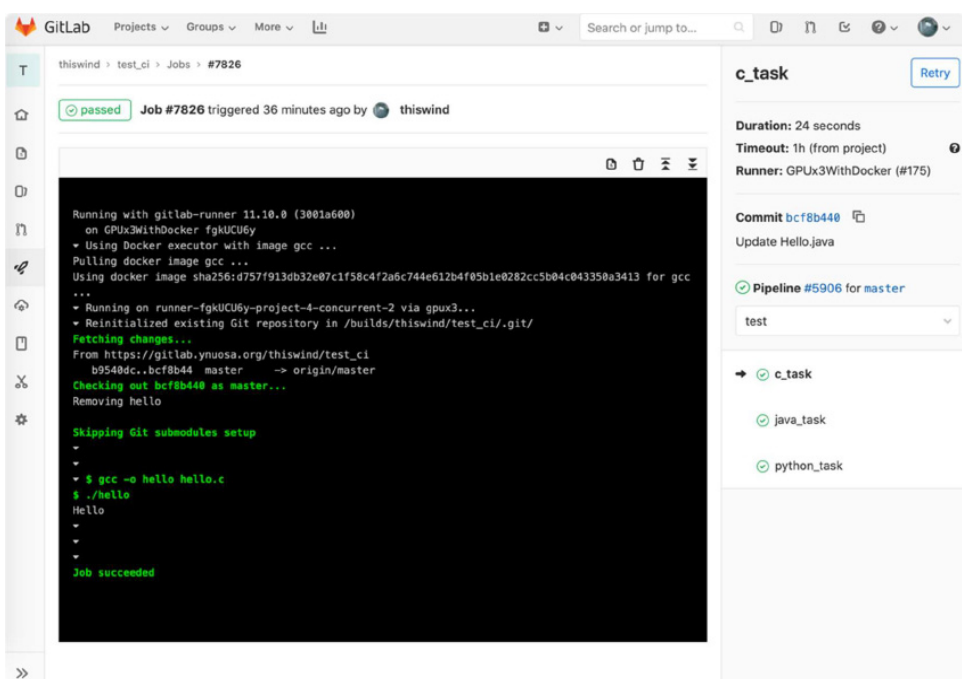

Figure 4 The execution result of the code in Figure 3

Through the platform, teachers can accurately understand the programming situation of each student, and the experimental class is not limited by time and space. The code and cases are increasingly rich and perfect, and students only have computers and networks, they can use "zero installation" mode. Students can submit code in real time and carry out automatic version management, and know whether the code is correct or not, high or low efficiency, class ranking. The platform can greatly improve the enthusiasm of students' programming.

## 3 Carry out programming ability certification

### 3.1 Introduction to programming ability certification in China

At present, China's well-known certification examinations for assessing programming ability mainly include the Certified Software Professional (CSP) sponsored by China Computer Federation [6] and the Programming Ability Test (PAT) sponsored by Zhejiang University[7]. The comparison is shown in **Table 3**.

Table 3 Comparison of CSP and PAT

| | Certified Software Professional (CSP) | Programming Ability Test (PAT) |
|---|---|---|
| Organizer | China Computer Federation(CCF) | Zhejiang University |
| Supported languages | C/C++、Java、Python | C/C++、Java |
| History | Started in 2014 | |
| Examination fee | Non-CCF members: RMB 300, CCF members: RMB180 | RMB 200 |
| Whether to grade | Only one set of questions per exam | Each exam is divided into top-level, first-level, and second-level exams, each of which has a different difficulty |
| Total test score | 500 points | 100 points |
| Exam duration | 4 hours | 3 hours |
| Examination time | Three times a year: March, September, December | |
| Last attendance (December 2019) | 11,395 | 2800 |
| Number of participating schools (as of December 2019) | 200+ | 61 |
| Number of examination sites (as of December 2019) | 126 | 63 |
| Test report information | Student information, programming language, scores and rankings, certificate verification code, etc | |

Both certification examinations focus on investigating participants analytical, problem-solving, and practical programming skills, and scientific evaluation of computer programming talents. They are more reflective of students paper test methods than traditional computer technology and software professional and technical qualification (level) exams. Practical hands-on ability. Both types of certification exams have their own advantages and disadvantages, both of which are good domestic program design certification exams.

Both certification examinations focus on the analysis, problem-solving and practical programming ability of the participants, and scientifically evaluate programming talents. Compared with traditional paper-based test, they can reflect the actual programming ability of the students. The both certification examinations have their own advantages and disadvantages. They are very good programming certification in China.

**3.2 Our methods and achievements**

In April 2017, we cooperated with the China Computer Federation (CCF), and CSP is introduced to provide certification sites for students and staff in Yunnan province. CSP provides a new direction for the teaching reform of the programming course, and provides a platform for students to test their programming ability, so as to continue to steadily improve the training quality of students.

In the revision of the 2017 computer science and technology training program, we added a "Programming Ability Test" curriculum in the comprehensive practice module. The course requires that each student's CSP score should reach the basic qualification line, or student pass China's computer software professional and technical qualifications examinations. Through this course, we can guarantee that the graduates can reach a certain level in the programming ability. In addition, CSP can also expand the employment channels of students. Hundreds of students with outstanding CSP score have been invited by more than ten famous IT enterprises that have cooperated with CCF. They are invited to participate in the recruitment and internship, which provides a good foundation for employment.

There are five questions in each CSP examination, and the difficulty increases in order. Students' basic language ability, algorithm ability, data structure organization ability, and model building ability are examined. The certification scores and specific correspondence are shown in **Table 4**.

Table 4 Correspondence between CSP scores and basic software capabilities [1,8]

| CSP score | Corresponding ability |
| --- | --- |
| 400-500 points | Divergent Algorithm Programming |
| 300-400 points | Ability to analyze and solve complex problems |
| 200-300 points | Structure organization ability, model building ability |
| 100-200 points | Basic language skills, simple algorithmic skills |

As of December 2019, we have conducted eight CSP certification examinations, with more than 1,400 students enrolled, and more than 1,100 students actually participated. The statistics data are shown in **Table 5**. Most of the students who take part in the examination come from the computer science and technology, software engineering, network engineering, Internet of things engineering, intelligent science and technology, information security, digital media technology of Yunnan University. A small number of students majoring in electronics, mathematics and physics also apply for the examination.

In the first two exams in 2017, the total average scores and the average scores of undergraduates in computer science and technology were not ideal because they were unfamiliar with the exam methods and question types. We then carried out the construction and reform of programming curriculums in the undergraduate curriculums of computer science and technology.

In order to test the effect of our reform, we extract the undergraduates majoring in computer science and technology from the registered students to compare after each examination. In the 2018 and 2019 certification, most students in this major can achieve CSP certification scores of 100-250 points. In the last six certifications, the average score of undergraduates in this major has exceeded the national average. The average score of the last two certifications has increased by more than 70 points compared with the average score in 2017, and the students' ability in programming has been significantly enhanced. Yunnan University has achieved outstanding results in organizing CCF CSP certification examination. In May 2018, it became one of the 39 CCF CSP authorized certification units. In July 2018 and July 2019, it was awarded excellent CSP certification units.

**Table 5** Comparison and analysis of CSP certification scores at Yunnan University

| Certification time | Number of certifications | Number of students | National average score | Average score of all participating students | Highest score | Average score of computer science and technology undergraduate |
|---|---|---|---|---|---|---|
| September 2017 | 11th CSP | 138 | 125 | 71 | 330 | 92.3 |
| December 2017 | 12th CSP | 144 | 136 | 105.5 | 255 | 94.1 |
| March 2018 | 13th CSP | 88 | 153 | 125.1 | 330 | **143.3** |
| September 2018 | 14th CSP | 100 | 144 | 103.5 | 300 | **148.5** |
| December 2018 | 15th CSP | 163 | 120 | 102.7 | 250 | **135.7** |
| March 2019 | 16th CSP | 110 | 106 | 95.1 | 300 | **127.8** |
| September 2019 | 17th CSP | 166 | 129 | 119.5 | 260 | **164.8** |
| December 2019 | 18th CSP | 202 | 131 | 96.6 | 310 | **163.5** |

## 4 Organize Programming Competitions

In order to better engage students' enthusiasm for programming, and to support the development of the school's second classroom, we select excellent students to participate in all kinds of programming competitions through CSP certification and school's programming competition. We can promote practice and learning by these competitions.

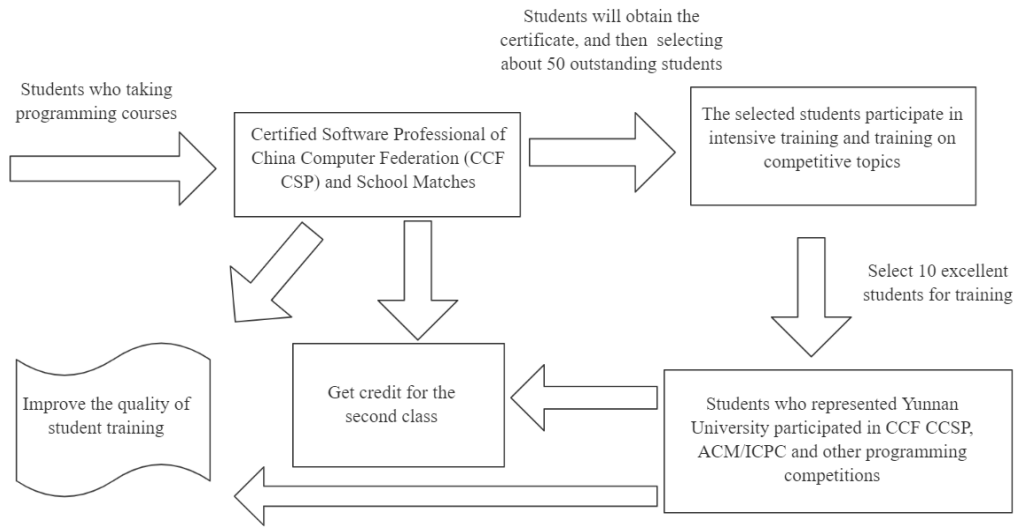

**Figure 5** The process of selecting programming competition students through CSP certification

Our main methods are as follows:

(1) Special funds are set up at our university to support programming competitions. In 2019, in order to better promote the construction of first-class undergraduate majors, Yunnan University set up a competition support project, and our programming competition project was approved.

(2) Organize programming competition of Yunnan University. During the science and technology festival of Yunnan University, we hosted the programming competition of Yunnan University, which is divided into preliminary competition and final competition. The number of participants in the preliminary competition has reached more than 300, and the contestant came from 6 colleges and 16 majors. 60 of them have been selected to participate in the final competition. They received certificates, bonuses, and second class credit awards.

(3) Actively organize to participate in all kinds of competitions and achieve excellent results. Through the school competition and CSP certification, we selected a group of students to participate in the CCF Collegiate Computer Systems & Programming Contest (CCSP), the national software and information technology professional competition of Blue Bridge Cup and other competitions, and Yunnan University students achieved excellent results. In May 2019, we won the champion and the third place (2 gold medals) of CCF CCSP in Southwest China. In October 2019, we won the silver medal of the CCF CCSP finals, which is the best performance of Yunnan Province in this competition.

## 5 Conclusion

As a practical engineering major, programming ability is one of the most important abilities of computer students. Taking Yunnan University as an example, this paper summarizes our explorations and practices. These reforms have achieved good results.

According to the needs of local economic and social development in Yunnan Province, based on the actual situation of teachers, students, resources and other aspects of Yunnan University, aiming at the construction of computer science and technology talents training, this paper discusses the summary of our work in improving the ability of programming, which can provide reference for similar universities. With the advancement of the construction of "New Engineering" and the implementation of the "Double 10000 plan", the requirements of computer skills in the era of artificial intelligence 2.0 have been continuously improved. How to further innovate the program design ability training mode, strengthen the engineering literacy training based on learning, and organically combine various teaching modes are the work we will carry out. We will carry out teaching reforms in natural language processing [9-10], image processing [11], AI-based education [12], and deep learning [13].